\documentclass[prd,amsmath,amssymb,preprint]{revtex4}
\usepackage{graphicx}
\usepackage[english]{babel}
\usepackage{feynmf}
\usepackage{dcolumn}
\usepackage{bm}
\usepackage{longtable}
\usepackage{pdflscape}
\begin{document}
\title{Isotope shifts of the 2p$_{3/2}$-2p$_{1/2}$ transition in B-like ions}
\author{N.~A.~Zubova$^{1,2}$, A.~V.~Malyshev$^{1,2}$,  I.~I.~Tupitsyn$^{1}$, V.~M.~Shabaev$^{1}$, Y.~S.~Kozhedub$^{1}$, G.~Plunien$^{3}$, C.~Brandau$^{4,5,6}$, and Th.~St\"ohlker$^{4,7,8}$}
\affiliation{$^1$Department of Physics, St. Petersburg State University, 7/9 Universitetskaya nab., St.~Petersburg
199034, Russia \\
$^2$SSC RF ITEP of NRC ``Kurchatov Institute'', Bolshaya Cheremushkinskaya 25, Moscow, 117218, Russia\\
$^3$Institut f\"ur Theoretische Physik, TU Dresden, Mommsenstrasse 13, Dresden, D-01062, Germany\\
$^4$ GSI Helmholtzzentrum f\"ur Schwerionenforschung GmbH, D-64291 
Darmstadt, Germany\\
$^5$  ExtreMe Matter Institute EMMI and Research Division, GSI 
Helmholtzzentrum f\"ur Schwerionenforschung, D-64291 Darmstadt, Germany\\
$^6$ Institut f\"ur Atom- und Molek\"ulphysik, Justus-Liebig-University Giessen, 
Leihgesterner Weg 217, D-35392 Giessen, Germany\\
$^7$ Helmholtz-Institut Jena, D-07743 Jena, Germany\\
$^8$ Institut f\"ur Optik und Quantenelektronik, 
Friedrich-Schiller-Universit\"at Jena, D-07743 Jena, Germany}
\date{\today}

\begin{abstract}
Isotope shifts of the 2p$_{3/2}$-2p$_{1/2}$ transition in B-like ions are evaluated 
for a wide range of the nuclear 
charge number: $Z=8-92$. The calculations of the relativistic nuclear recoil and nuclear size effects are performed using a large scale configuration-interaction Dirac-Fock-Sturm method. The corresponding QED corrections are also taken into account. The results of the calculations are compared with the theoretical values obtained with other methods. The accuracy of the isotope shifts of the $2p_{3/2}-2p_{1/2}$ transition in B-like ions is significantly improved.
\end{abstract}

\pacs{Valid PACS appear here}
\maketitle
\section{Introduction}
First measurements to isolate the isotopic variation of nuclear effects in the binding energies in few-electron highly charged ions were performed in Refs. 
\cite{Elliot_1996, Elliot_1998, Schuch_2005}. The most precise to-date measurements of the isotope shifts were carried out for B-like argon \cite{Orts_2006} and Li-like neodymium \cite{Brandau_2008}. These experiments allowed first tests of the relativistic theory of the mass shift with middle- and high-Z systems for the first time. 

From the theoretical side the first evaluation of the isotope shifts in boronlike argon was performed in Ref. {\cite{Tupitsyn_2003,Orts_2006}}. 
Later, systematic calculations of the relativistic nuclear recoil effect were performed by {\it{Kozhedub et al.}} \cite{Kozhedub_2010}, who used a large-scale configuration interaction Dirac-Fock-Sturm (CI-DFS) method, by {\it Li et al.}  \cite{Li_2012}, who employed the multiconfiguration Dirac-Fock (MCDF) method, and also in our recent work \cite{Zubova_2014}, where the perturbation theory calculations and the CI-DFS method were combined. The CI-DFS and MCDF methods are simpler in using compared to the perturbative methods, but they show a 
rather poor convergence in calculations of the specific mass shift.
The results of the calculations of the relativistic nuclear recoil effect obtained by the CI-DFS method \cite{Kozhedub_2010} were confirmed by the perturbative calculations of the interelectronic-interaction corrections to the mass shifts \cite{Zubova_2014}. In Refs. \cite{Orts_2006,Brandau_2008,Tupitsyn_2003} it was found that quantum electrodynamics (QED) corrections to the isotope shifts are of the same order of magnitude as the experimental uncertainties. Therefore, the high-precision calculations of the isotope shifts have to take into account the QED corrections.

The main goal of this paper is the high-precision evaluation of the isotope shifts for the $2p_{3/2}-2p_{1/2}$ transition in highly charged boronlike ions. Due to the relativistic origin of the $2p_{3/2}-2p_{1/2}$ splitting, the study of this transition provides a unique opportunity for tests of the relativistic and QED nuclear recoil effects in the nonperturbative regime. This is due to the fact that,  
in contrast to light atoms, the calculations of highly charged ions have to be performed to all orders in the parameter $\alpha Z$ ($\alpha$ is the fine structure constant and $Z$ is the nuclear charge number). 
It is expected that with new FAIR facilities \cite{fair} the experimental accuracy of the isotope shift measurements with highly charged Li- and B-like ions will be improved by an order of magnitude. To meet this accuracy, the high-precision calculations of the relativistic and QED contributions to the isotope shifts must be performed. In our calculations, along with the main contributions (the relativistic nuclear recoil and the nuclear size effect) 
the related QED corrections are also evaluated.
Moreover, the calculations of the QED corrections to the nuclear size contribution include the screening effects, which for the $2p_{3/2}-2p_{1/2}$ splitting in highly charged B-like ions are comparable with the first-order corrections. The calculations are performed for the nuclear charge number in a wide range: $Z=8-92$.

The relativistic units ($\hbar=c=1$) are used in the paper.

\section{Relativistic nuclear recoil effect}
To the lowest order in $m/M$, the relativistic nuclear recoil (mass shift) Hamiltonian $H_M$ within the Breit approximation is given by \cite{Shabaev_1985, Shabaev_1988, Palmer_1987, Shabaev_1998r}:
\begin{eqnarray}   
\label{rec1}
 H_M &=& \frac{1}{2M}\sum_{i,k}\Bigl[
{\vec{p}_i}\cdot {\vec{p}_k} -\frac{\alpha Z}{r_i} \Bigl[
{\vec{\alpha}_i}+\frac{({\vec{\alpha}_i} \cdot {\vec{r}_i}){\vec{r}_i}}{{r_i}^2}
\Bigr]\cdot {\vec{p}_k}\Bigr],
\end{eqnarray}
 where the indices $i$ and $k$ numerate the atomic electrons, 
$\vec{p}$ is the momentum operator, $\vec{\alpha}$ are the Dirac
 matrices.

The operator (\ref{rec1}) can be represented by a sum: 
\begin{equation}
\label{rec2}
H_M = H_{\rm{NMS}}+H_{\rm{RNMS}}+H_{\rm{SMS}}+H_{\rm{RSMS}},
\end{equation}
where 
\begin{equation}   
\label{rec_NMS}
 H_{\rm{NMS}} = \frac{1}{2M}\sum_{i}{\vec{p}_i}^2\, 
\end{equation}
is the normal mass shift (NMS) operator,
\begin{equation}   
\label{rec_RNMS}
H_{\rm{RNMS}} = -\frac{1}{2M}\sum_{i}\frac{\alpha Z}{r_i}\Bigl[\vec{\alpha}_i+\frac{(\vec{\alpha}_i\cdot\vec{r}_i)\vec{r}_i}{r_i^2}\Bigr]\cdot\vec{p}_i\, 
 \end{equation}
is the relativistic normal mass shift (RNMS) operator,
\begin{equation}   
\label{rec_SMS}
 H_{\rm{SMS}} = \frac{1}{2M}\sum_{i\neq k}{\vec{p}_i\cdot \vec{p}_k}\, 
\end{equation}
is the specific mass shift (SMS) operator, and
\begin{equation}   
\label{rec_RSMS}
H_{\rm{RSMS}} = -\frac{1}{2M}\sum_{i\neq k}\frac{\alpha Z}{r_i}\Bigl[\vec{\alpha}_i+\frac{(\vec
{\alpha}_i\cdot\vec{r}_i)\vec{r}_i}{r_i^2}\Bigr]\cdot\vec{p}_k\,
\end{equation}
is the relativistic specific  mass shift (RSMS) operator.

In the present paper, we evaluate the relativistic nuclear recoil contribution within the Breit approximation  to all orders in $1/Z$. The calculation is carried out by averaging the operator (\ref{rec2}) with the eigenvectors of the Dirac-Coulomb-Breit (DCB) Hamiltonian:
\begin{equation}
\Delta E=\langle \psi |H_{{M}}| \psi \rangle,
\end{equation} 
where the wave function $| \psi \rangle$ is evaluated using
the configuration-interaction Dirac-Fock-Sturm method \cite{Tupitsyn_2003} for an extended nucleus.
Details of the calculations are presented in Sec. \ref{sec4}.
An independent evaluation of the non-QED mass shifts based on the multiconfiguration Dirac-Fock method was presented in Ref. 
\cite{Naze_2014}.   
For B-like argon the results of this calculation agree with those from Ref. {\cite{Tupitsyn_2003}}. In the present paper we extend the calculations of Ref. {\cite{Tupitsyn_2003}} to B-like ions in the range $Z=8-92$. 
The obtained non-QED results are combined with the corresponding QED contributions evaluated to the zeroth order in $1/Z$ to get the most accurate theoretical data for the mass shifts in highly charged B-like ions.    

In Ref. \cite{Aleksandrov_2015}, the nuclear size correction to the recoil operator (\ref{rec1}) was studied for H-like ions. It was found that for heavy ions this correction can amount to about 20 \% of the total nuclear size contribution to the recoil effect. We estimate that this correction, combined with the related QED nuclear size correction, should be within the total uncertainties presented in this paper. 

\section{Finite nuclear size effect}
The finite nuclear size effect (the so-called field shift) is caused by the difference in the  nuclear charge distibution of the isotopes. The main contribution to the field shift can be calculated in the framework of the Dirac-Coulomb-Breit Hamiltonian. 
The nuclear charge distribution is usually approximated by the spherically-symmetric Fermi model:
\begin{equation}
\label{rho}
\rho(r,R)=\frac{N}{1+{\rm{exp}}[(r-c)/a]},
\end{equation}
where the parameter $a$ is generally fixed to be $a=2.3/(4{\rm ln}3)$ fm and the parameters $N$ and $c$ are determined using the given value of the root-mean-square ($\rm{rms}$) nuclear charge radius $R=\langle r^2 \rangle^{1/2}$ and the normalization condition: $\int{d\vec{r} \rho({r},R)}=1$. 
The potential induced by the nuclear charge distribution $\rho(r,R)$ is defined as 
\begin{equation}
\label{Vn}
 V_{N}(r,R)= -4\pi \alpha Z \int\limits_{0}^{\infty} {dr' r'^2 \rho (r',R) \frac{1}{r_{>}}},
\end{equation}
where $r_{>}={\rm{{max}}}(r,r')$. 
Since the finite nuclear size effect is mainly determined by the $\rm{rms}$ nuclear charge radius (see, e.g., Ref. \cite{Shabaev_1993}), the energy difference between two isotopes can be approximated as
\begin{equation}
\label{FS_1}
\delta E_{FS} = {F}\delta \langle r^2 \rangle,
\end{equation}
where $F$ is the field shift factor and $\delta \langle r^2 \rangle$ is the 
mean-square charge radius difference. 
In accordance with this definition, in the present paper the $F$-factor is evaluated by
\begin{equation}
\label{FS_2}
F= \langle \psi \mid \sum_{i} \frac{dV_{N}(r_{i},R)}{d\langle r^2 \rangle} \mid  \psi \rangle,
\end{equation}
where $\psi$ is the wave function of the state under consideration and the index $i$ runs over all atomic electrons. 
These calculations, being performed by the CI method in the basis of the Dirac-Fock-Sturm orbitals, are compared with the corresponding MCDF calculations of Ref. \cite{Naze_2014}, where the $F$-factor was approximated by

\begin{equation}
\label{FS_3}
F=\frac{2 \pi}{3} \alpha Z {\mid \psi(0) \mid}^2.
\end{equation}

In addition, the QED corrections to the field shifts have been evaluated. The calculations have been performed by perturbation theory including the second-order screening effects in accordance with the technique presented in Refs. \cite{Malyshev_2014,Malyshev_2015}.

\section{Results and discussion}
\label{sec4} 
The nuclear recoil contributions can be represented in terms of the $K$-factor,
\begin{equation}
\Delta E=\frac{K}{M}.
\end{equation}
Then, the isotope mass shift is determinated by
\begin{equation}
\delta E_{\rm{MS}}=\frac{K}{M_1}-\frac{K}{M_2}=-\frac{\delta M}{M_1 M_2}K,
\end{equation}
where $\delta M=M_1-M_2$ is the nuclear mass difference. 

To calculate the relativistic nuclear recoil contributions, we use the large-scale
configuration-interaction method with the basis of the Dirac-Fock-Sturm orbitals.
The excited configurations are obtained from the basic configuration via a single, double
and triple excitations of electrons. The accuracy of the
calculations is defined by a stability of the results with respect to a variation
of the basis size. In the present paper we use two different 
sets of the electron orbitals. In our notations, the middle basis is the basis which 
includes all orbitals with the excitations up to $(10s\,\,10p\,\,10d\,\,10f\,\,10g)$ shells. 
The large basis includes the excitations up to
$(15s\,\,15p\,\,15d\,\,12f\,\,12g\,\,12f)$.

To estimate the quality of the bases used we have
performed the following test. First of all, we have extracted the contribution of the order $1/Z$
from the total value of the relativistic nuclear recoil correction (7) 
obtained within the CI-DFS calculations.
To this end, we have used the procedure, which was described, for example, in Ref. \cite{Zubova_2014} (see
also Refs. \cite{Glazov_2004,Glazov_2006,Artemyev_2007,Kozhedub_2010}, where a similar method 
was applied to separate the interelectronic-interaction contributions
of the different orders in $1/Z$). In accordance with this 
 procedure, the DCB Hamiltonian is represented
as a sum of two parts:
\begin{eqnarray}
\label{H_1}
H &=& H_0 + \lambda V, \\
\label{H_l}
H_0 &=& \sum_i \left[h_{\rm D}^{(i)} + V_{\rm scr}^{(i)}\right], \\
\label{H_2}
V &=& \sum_{i<j}\, V(i,j) - \sum_i\, V_{\rm scr}^{(i)}, 
\end{eqnarray}
\begin{eqnarray}
V(i,j)=V_{\rm{C}}(i,j)+V_{\rm{B}}(i,j)
=\frac{\alpha}{r_{ij}}
-\alpha\Bigr[
\frac{\vec \alpha_i\cdot{\vec \alpha_j }}
{r_{ij}}+\frac{1}{2}({\vec \nabla_i} \cdot {\vec \alpha_i})({\vec \nabla_j} 
\cdot {\vec \alpha_j})r_{ij}\Bigr].
\end{eqnarray}
Here $H_0$ is the unperturbed Hamiltonian ($h_{\rm D}$ is the one-electron Dirac Hamiltonian),
$V$ describes the perturbation by the Coulomb and Breit interelectronic interaction, $\lambda$ is a free parameter 
with the physical value equal to $1$, and the summation goes over  all electrons of the system.
For each electron we have added some local screening potential $V_{\rm scr}$ to the unperturbed 
Hamiltonian $H_0$. To avoid the double counting, the corresponding contribution has to be subtracted 
from the interaction part $V$. 

For small $\lambda$, the nuclear recoil contribution can be expanded in powers of  $\lambda$:
\begin{equation}
E_{\rm{MS}}(\lambda)=E^{(0)}_{{\rm{MS}}}+E^{(1)}_{{\rm{MS}}} \lambda +\sum \limits_{k=2}^
{\infty}{E^{(k)}_{{\rm{MS}}} \lambda^k},
\end{equation}
where
\begin{equation}
\label{Ek}
E^{(k)}_{{\rm{MS}}}=\frac{1}{k!}\frac{d^k}{d\lambda^k}E_{{MS}}(\lambda)_
{|{\lambda=0}}.
\end{equation}
It is easy to see that the coefficient $E_{\rm{MS}}^{(1)}$ corresponds to the contribution of the order $1/Z$
to the total relativistic nuclear recoil correction (7).  
Calculating the derivatives 
we have evaluated the first-order  corrections
to the $2p_{3/2}-2p_{1/2}$ transition energies in B-like oxygen, fluorine, and uranium. 
The results of the calculations with the middle and large bases are presented in the first
and second columns of Table \ref{tab1}, respectively. 

On the other hand, the first-order relativistic nuclear recoil correction
can be evaluated by the standard perturbation theory. For a non-degenerate state 
$a$ the corresponding
correction may be expressed in the following form:
\begin{eqnarray}
\label{E_1}
E_{\rm{MS}}^{(1)} &=&2\sum_{n\neq a} \frac{\langle a | H_M | n
\rangle \langle n | V | a \rangle}{\varepsilon_a-\varepsilon_n}.
\end{eqnarray}
Since we have introduced
the screening potential into the zeroth-order Hamiltonian (\ref{H_l}), we can avoid the quasidegeneracy between $1s^2 2s^2 2p_{3/2}$ and $1s^2 (2p_{1/2})^2 2p_{3/2}$
states that takes place if the pure Coulomb field is employed in the zeroth-order approximation. 
In the specific calculations we use the local Dirac-Fock (LDF) screening potential \cite{Shabaev_2005}.

From Eq. (\ref{rec1}) one can see that the relativistic recoil operator $H_M$ mixes 
the states with the different values of the orbital quantum number $l$, but $l$ can not differ more than by unity. All electrons in the states
under consideration ($1s^2 2s^2 2p_{1/2}$ and $1s^2 2s^2 2p_{3/2}$) have $l$ =$0$ or $l=1$.
Therefore, without loss of generality we can restrict the summation over the spectrum in Eq. (\ref{E_1}) to the summation
over the $s$, $p$ and $d$ states. 
We have performed the calculation of the relativistic nuclear recoil
correction for the $2p_{3/2}-2p_{1/2}$ transition energies to the first order in $1/Z$ using Eq. (\ref{E_1}).
For this aim we have employed the extra large basis with the number of the orbitals doubled: 
$(30s\,\,30p\,\,30d)$. The results of this calculation are given in the last column of Table \ref{tab1}.

From Table \ref{tab1} it is seen that for light ions the 
contributions of the order $1/Z$ to the recoil effect obtained with the middle and large bases
differ significantly. It is clear that
the middle basis is not sufficient to perform the calculations for low- and middle-$Z$ ions.
At the same time, the results of the calculations of the first-order correction by the CI-DFS method with the large basis 
and by the direct summation over the spectrum within the standard perturbation theory
are in a good agreement with each other. This agreement shows that
the total values $E_{\rm{MS}}$ obtained in the large basis and
the numerical derivatives in Eq. (\ref{Ek}) are evaluated with a good accuracy. Therefore, the final calculations will be
performed mainly with the usage of the large basis.  

In Table \ref{tab2} the contributions to 
$K_{\rm{NMS}}$, $K_{\rm{SMS}}$, 
$K_{\rm{RNMS}}$, and $K_{\rm{RSMS}}$ for the $2p_{3/2}-2p_{1/2}$ transition in boron-like ions 
in the range $Z=8-92$ are presented. To find these corrections we have averaged the
relativistic nuclear recoil operators (\ref{rec_NMS})-(\ref{rec_RSMS}) with the CI-DFS functions. The set of the configuration state functions (CSFs) was obtained using the restricted active space method with the single and double exitations only. Here, the basis of the virtual orbitals was chosen in the middle form. Table \ref{tab3} demonstrates the role of the triple excitations. Comparing Table \ref{tab2} and \ref{tab3}, one can see that taking into account the triple excitations changes the values of $K_{\rm{NMS}}$, $K_{\rm{SMS}}$,
$K_{\rm{RNMS}}$, and $K_{\rm{RSMS}}$ contributions slightly.

Further, in Table \ref{tab4} we present the results of the calculations of the individual contributions to the total values of the non-QED mass shift, which were obtained with the usage of the large basis.  
It should be noted that the
direct calculations including the triple excitations turned out to be too time
consuming. For this reason,
the triple excitation contribution $\Delta_{\rm {triple}}$ was obtained as the 
difference between the total values 
$K_{MS}$ from Tables \ref{tab3} and \ref{tab2}. 
This approach to the calculation of the 
triple excitation contribution was confirmed by the full CI-DFS calculation for the $2p_{3/2}-2p_{1/2}$ transition in B-like oxygen ($K=-$0.0979$\times 10^{2}$ GHz$\cdot$amu) and in B-like uranium ($K=-$136.8 $\times 10^{4}$ GHz$\cdot$amu). Generally, our results are in a reasonable agreement with the results of the MCDF calculations of Ref. {\cite{Naze_2014}}.
However, there is some discrepancy for the lightest ions (about 8 $\%$ for oxygen and fluorine ions). The reason for this discrepancy is unclear to us.

Finally, we should take into account the nuclear recoil effects beyond the Breit approximation (the so-called $\rm{QED}$ nuclear recoil terms). The calculations of the QED terms for highly charged ions to the zeroth order in $1/Z$ were performed in Refs. {\cite{Kozhedub_2010, Artemyev_1995, Artemyev2_1995,Shabaev_1998_2, Shabaev_1999, Adkins_2007}. 
In the present paper, we have recalculated these corrections and found some misprints in Table 2 of Ref. {\cite{Artemyev_1995}}, where the two-electron contributions for the ${(1s)}^2 2p_{3/2}$ state were presented. Namely, the values in the fifth column of that table, which are supposed to be equal to the sum of the values given in the second, third, and fourth columns, are incorrect. The correct values are listed in Table \ref{tab5} of the present paper. This table displays the total two-electron mass-shift contributions of the zeroth order in $1/Z$, which are expressed in terms of the function $Q(\alpha Z)$:
\begin{equation}
\label{2QED}
\Delta E=-\frac{m^2}{M}\frac{2^9}{3^8}{(\alpha Z)}^2 Q(\alpha Z).
\end{equation}  
The calculations are performed for both point and extended nuclei. In the extended nucleus case, the Fermi model of the nuclear charge distribution was used for $Z \ge 20$ and the model of the homogeneously charged sphere otherwise. The two-electron QED corrections are obtained by subtracting the corresponding contributions in the Breit approximation.

In Table \ref{tab6} we present the total values of the mass shifts in the range $Z=8-92$. The total values of the non-QED mass shifts are defined according to Table \ref{tab4}. Namely, the non-QED contributions to the nuclear recoil effect were evaluated with the usage of the CI-DFS method, taking into account the single, double, and triple excitations. The QED corrections have been evaluated in the independent-electron approximation. The calculations have been performed for both the Coulomb potential and the effective potential (the extended Furry picture). As the effective potential we have used the LDF potential. 

The uncertainty was estimated as a quadratic sum of the uncertainty due to the CI-DFS calculations, the uncertainty obtained by changing the potential from the Coulomb to 
the local-Dirac-Fock in the QED contribution of the zeroth order in $1/Z$, and 
the uncertainty due to uncalculated $\rm{QED}$ contributions of the first order in $1/Z$. The latter one was evaluated as the $\rm{QED}_{Coul}$ contribution 
of the zeroth order in $1/Z$ multiplied with a factor $2/Z$ ( in the same way as in Ref. \cite{Zubova_2014}). For $Z \ge 60$, an uncertainty due to nuclear size corrections to the recoil operator (including the QED part) has been also added.


To calculate the field shift constants within the Breit approximation we have used the CI-DFS method. Table \ref{tab7} presents the non-QED $F$-factor, obtained 
according to Eq. 
(\ref{FS_2}). In the third column we give the DF results, while the fourth column presents the results of the CI-DFS calculations, including the Breit electron-interaction correction. 
The results of Ref. \cite{Naze_2014}, where the formula (\ref{FS_3}) was employed, are presented in the last column. It can be seen that the calculations by formula (\ref{FS_3}) have 
 a rather poor accuracy for heavy ions. In case of B-like molybdenum the discrepancy between the results obtained with the  
equations (\ref{FS_2}) and (\ref{FS_3})  amounts to by about 7 \%. Some discrepancy with the results of Ref. {\cite{Naze_2014}} for low-Z ions can be explained by a rather strong cancellation of the significant digits in the $2p_{3/2}-2p_{1/2}$ energy difference, since in Ref. {\cite{Naze_2014}} the $F$-constants are presented only for the $1s^2 2s^2 2p_{1/2}$ and $1s^2 2s^2 2p_{3/2}$ states, and not for the differences. The uncertainty due to the CI-DFS calculations was estimated as in the mass shift case.

In Fig. \ref{FS_graph1}, we present the normalized FS constant for the $2p_{3/2}-2p_{1/2}$ transition, which is determinated as 
$\Delta F/F_0$, where $\Delta F=F_{2p_{3/2}}-F_{2p_{1/2}}$  and $F_0$ is the field shift factor for the $2p_{1/2}$ state of the corresponding H-like ion, obtained using analytical formulas from Ref. \cite{Shabaev_1992}.
In accordance with Ref. \cite{Shabaev_1992},
\begin{equation}
\label{F0}
F_0=\frac{\gamma \Delta E_{2p_{1/2}}}{R^2},
\end{equation}
where
\begin{equation}
\label{FNS}
\Delta E_{2p_{1/2}}=
\frac{Z^4 \alpha^2 (n^2-1)}{40n^3}[1+(Z\alpha)f_{2p_{1/2}}(\alpha Z)] \Bigr( \frac{2Z\alpha R}
{n \lambda_c} \Bigr)^{2\gamma}
\end{equation}
and
\begin{equation}
f_{2p_{1/2}}(Z\alpha)=1.615+4.319(Z\alpha)-9.152(Z\alpha)^2+11.87(Z\alpha)^3.
\end{equation}
In Fig. \ref{FS_graph1}, the dotted line indicates the results for the H-like ions obtained using Eq. (\ref{FS_2}), the dashed line shows the results of the DF calculations using Eq. (\ref{FS_2}), the dashed-dotted line stands for the 
CI-DFS calculations using the approximate formula (\ref{FS_3}), and the solid line represents the results of the CI-DFS calculations using Eq. (\ref{FS_2}). We observe that for low-Z ions the sign of $\Delta F$ becomes positive. This is due to different 
$(1s)^2 (2s)^2$ core polarizations by the 
$2p_{3/2}$ and $2p_{1/2}$ states.
We note also that this effect is weaker in the CI-DFS calculations (solid line) in comparison with the DF calculations (dashed line) because of the admixing of additional configurations to the main configuration.

Table \ref{tab8} presents the QED corrections to the field-shift $F$-constant for the 
$2p_{1/2}-2s$ and $2p_{3/2}-2s$ transitions in high-Z Li-like ions, and also for the $2p_{3/2}-2p_{1/2}$ transition in high-Z B-like ions. The {\textit {ab initio}} calculations of the QED corrections to the finite nuclear size effect have been considered by perturbation theory in the first two orders \cite{Malyshev_2014}. In the case of the $2p_{3/2}-2p_{1/2}$ transition in B-like ions it turned to be very essential to take into account the two-electron self-energy and vacuum polarization corrections.
The one-electron finite nuclear size QED corrections for the $2p_{1/2}$ and $2p_{3/2}$ states are significantly smaller than the corresponding corrections for the $s$-states. For this reason, although the screening contributions are generally suppressed by the factor $1/Z$ compared to the one-electron  contributions, the interaction with the $1s^2 2s^2$ core makes the two-electron finite nuclear size corrections comparable with the contribution of the leading order. Besides, owing to the strong cancellation between the self-energy and vacuum polarization finite nuclear size effects on the $2p_{3/2}-2p_{1/2}$ transition, for high-Z ions it is also important to consider the contribution from the energy dependence of the interelectronic interaction operator, that is beyond the Breit approximation.

In our previous work \cite{Zubova_2014} the QED corrections to the field-shift $F$-constant for the
$2p_{1/2}-2s$ and $2p_{3/2}-2s$ transitions in Li-like ions were taken into account using the approximate analytical formulas for H-like ions from Refs. \cite{Milstein_2004, Yerokhin_2011}. This was done by multiplying the $s$-state QED correction factor 
${\Delta}_{s}$ with the total nuclear size contribution to the corresponding transition energy. 
The values of the QED corrections obtained in this way are also presented in Table \ref{tab8}.
We note that for the $2p_{3/2}-2p_{1/2}$ transition in high-Z B-like ions the QED corrections contribute on the level of the total uncertainty of the field-shift $F$-constants.

In Table \ref{tab9} we present the isotope shift of the $2p_{3/2}-2p_{1/2}$ transition in B-like argon with atomic numbers $A$=36 and $A$=40. The values of $\delta {\langle r^2 \rangle}^{1/2}$ are taken from Ref. {\cite{Angeli_2013}}. 
The relativistic nuclear recoil and finite nuclear size effects and 
the corresponding QED corrections are taken from Tables  \ref{tab6} and \ref{tab7}. 
The perfect agreement of the present theoretical value with that of Refs. \cite{Tupitsyn_2003,Orts_2006,Naze_2014} and with
the experiment \cite{Orts_2006} is observed. A small discrepancy of the non-QED part 
between the present work and Ref. \cite{Naze_2014} is within the uncertainty.

In Table \ref{tab10} we present the isotope shifts of the $2p_{3/2}-2p_{1/2}$ transition in B-like uranium for two pairs of  even-even isotopes, 
 $^{238}{\rm{{U}}^{87+}}-^{236}{\rm{{U}}^{87+}}$ and $^{238}{\rm{{U}}^{87+}}-^{234}{\rm{{U}}^{87+}}$. 
The nuclear polarization effect was evaluated using the results of Refs. \cite{Plunien_1991,Plunien_1995,Nefiodov_1996,Volotka_2014}. The nuclear deformation effect was calculated as in Ref. \cite{Kozhedub_2008}, using the experimental \cite{Bemis_1973} and theoretical \cite{Zumbro_1984} data for
the nuclear deformation parameters.


\section{Conclusion}
In this paper we have evaluated the isotope shifts of the $2p_{3/2}-2p_{1/2}$ transition energies in boron-like ions. 
The configuration-interaction method in the Dirac-Fock-Sturm basis was employed to calculate the relativistic nuclear recoil and the finite-nuclear size effects within the framework of the Dirac-Coulomb-Breit Hamiltonian. The obtained results are compared with the previous related calculations. The QED nuclear recoil corrections have been evaluated within the independent-electron approximation. The QED corrections to the field shift have been calculated by perturbation theory including the self-energy and vacuum-polarization contributions of the zeroth and first orders in $1/Z$. As the results, the most accurate theoretical predictions for the mass shifts and field shifts of the $2p_{3/2}-2p_{1/2}$ transition energies in boron-like ions have been obtained.

\clearpage
\begingroup
\squeezetable

\begin{table*}[H]
\caption{\label{tab1} Mass shift contributions in terms of the $K$-factor
(in units of 1000 GHz$\cdot$amu) evaluated to the first order in $1/Z$ for the $2p_{3/2}-2p_{1/2}$
transition in B-like oxygen, fluorine, and uranium. 
The calculations are performed in the middle, large, and extra large bases of the virtual orbitals.}
\begin{ruledtabular}
\begin{tabular}{ccccc}
Ion & middle (CI-DFS) &large (CI-DFS) & extra large (PT)\\ 
O$^{3+}$ & 0.4644$\times 10^{-2}$ & 0.5564$\times 10^{-2}$ & 0.5560$\times 10^{-2}$\\
F$^{4+}$ & 0.7929$\times 10^{-2}$ &  0.8447$\times 10^{-2}$ &0.8440$\times 10^{-2}$\\
U$^{87+}$ & 0.2641$\times 10^{2}$ & 0.2675$\times 10^{2}$ & 0.2674$\times 10^{2}$\\
\end{tabular}
\end{ruledtabular}
\end{table*}

\begin{table*}[H]
\caption{\label{tab2}Mass shift contributions in terms of the $K$-factor (in units of 1000 GHz$\cdot$amu) for the
$2p_{3/2}-2p_{1/2}$
transition in B-like ions. The calculations are performed in the middle basis $(10s\,\,10p\,\,10d\,\,10f\,\,10g)$ of the virtual orbitals, accounting only for the single and double excitations.}
\begin{ruledtabular}
\begin{tabular}{cccccccc}
Ions &  ${\langle r^2 \rangle}^{1/2}$ & NMS & SMS & RNMS & RSMS & Total \\
O$^{3+}$ & 2.6991 &-0.1996$\times 10^{-1}$ &0.1737$\times 10^{-1}$ & 0.1887$\times 10^{-1}$ & -0.2665$\times 10^{-1}$ & -0.1037$\times 10^{-1}$  \\
F$^{4+}$ &2.8976 & -0.3740$\times 10^{-1}$ & 0.3191$\times 10^{-1}$ & 0.3445$\times 10^{-1}$ & -0.4869$\times 10^{-1}$ & -0.1973$\times 10^{-1}$   \\
Ne$^{5+}$ & 3.0055& -0.6584$\times 10^{-1}$ &  0.5429$\times 10^{-1}$ & 0.5813$\times 10^{-1}$ & -0.8214$\times 10^{-1}$ & -0.3557$\times 10^{-1}$\\
Na$^{6+}$ &  2.9936 &-0.1070 &0.8613$\times 10^{-1}$ &  0.9207$\times 10^{-1}$ & -0.1301 & -0.5891$\times 10^{-1}$\\
Al$^{8+}$  & 3.0610 &-0.2443 & 0.1880 & 0.2019 & -0.2848 & -0.1392  \\
P$^{10+}$  & 3.1889 &-0.4848 & 0.3585 & 0.3893 & -0.5479 & -0.2850 \\
S$^{11+}$  & 3.2611 &-0.6569 & 0.4769 & 0.5216 & -0.7332 & -0.3916 \\
Cl$^{12+}$ & 3.3654 &-0.8717 & 0.6221 & 0.6855 & -0.9622 & -0.5263 \\
Ar$^{13+}$ & 3.4028 &-0.1136$\times 10^{1}$ & 0.7976 & 0.8859 & -0.1241$\times 10^{1}$ & -0.6939\\
K$^{14+}$  & 3.4349 &-0.1457$\times 10^{1}$ & 0.1008$\times 10^{1}$ & 0.1128$\times 10^{1}$ & -0.1578$\times 10^{1}$ & -0.8995  \\
Ca$^{15+}$ & 3.4776 & -0.1843$\times 10^{1}$ & 0.1256$\times 10^{1}$ & 0.1418$\times 10^{1}$ & -0.1980$\times 10^{1}$ & -0.1148$\times 10^{1}$\\
Sc$^{16+}$ & 3.4776& -0.2302$\times 10^{1}$ & 0.1549$\times 10^{1}$ & 0.1761$\times 10^{1}$ & -0.2455$\times 10^{1}$ & -0.1446$\times 10^{1}$\\
Ti$^{17+}$ & 3.5921& -0.2845$\times 10^{1}$ &  0.1891$\times 10^{1}$ & 0.2166$\times 10^{1}$ &  -0.3012$\times 10^{1}$ &  -0.1799$\times 10^{1}$\\
V$^{18+}$  & 3.6002&-0.3480$\times 10^{1}$ &  0.2288$\times 10^{1}$ &  0.2638$\times 10^{1}$ & -0.3660$\times 10^{1}$ & -0.2214$\times 10^{1}$ \\
Cr$^{19+}$ & 3.6452& -0.4218$\times 10^{1}$ & 0.2746$\times 10^{1}$ & 0.3186$\times 10^{1}$ & -0.4410$\times 10^{1}$ & -0.2696$\times 10^{1}$\\
Fe$^{21+}$  & 3.7377& -0.6053$\times 10^{1}$ &  0.3878$\times 10^{1}$ & 0.4545$\times 10^{1}$ & -0.6260$\times 10^{1}$ & -0.3890$\times 10^{1}$\\
Co$^{22+}$ & 3.7875 & -0.7174$\times 10^{1}$ & 0.4567$\times 10^{1}$ & 0.5375$                                
\times 10^{1}$ & -0.7384$\times 10^{1}$ & -0.4616$\times 10^{1}$\\
Cu$^{24+}$ & 3.8823& -0.9891$\times 10^{1}$ & 0.6244$\times 10^{1}$ & 0.7385$\times 10^{1}$ & -0.1009$\times 10^{2}$ &-0.6354$\times 10^{1}$\\
Zn$^{25+}$ &  3.9491&-0.1153$\times 10^{2}$ & 0.7247$\times 10^{1}$ & 0.8591$\times 10^{1}$ & -0.1171$\times 10^{2}$ &-0.7396$\times 10^{1}$ \\
Kr$^{31+}$& 4.1835& -0.2622$\times 10^{2}$ & 0.1651$\times 10^{2}$ &  0.1947$\times 10^{2}$ & -0.2611$\times 10^{2}$ & -0.1635$\times 10^{2}$\\
Mo$^{37+}$ & 4.3151 & -0.5279$\times 10^{2}$ & 0.3401$\times 10^{2}$ &  0.3927$\times 10^{2}$ & -0.5193$\times 10^{2}$ & -0.3144$\times 10^{2}$\\
Xe$^{49+}$ & 4.7964& -0.1695$\times 10^{3}$ & 0.1154$\times 10^{3}$ &  0.1280$\times 10^{3}$ & -0.1655$\times 10^{3}$ & -0.9161$\times 10^{2}$ \\
Nd$^{55+}$ & 4.9123&  -0.2813$\times 10^{3}$ & 0.1952$\times 10^{3}$ & 0.2149$                                
\times 10^{3}$ & -0.2752$\times 10^{3}$ & -0.1463$\times 10^{3}$\\
Yb$^{65+}$ & 5.0423& -0.6108$\times 10^{3}$ & 0.4295$\times 10^{3}$ &  0.4790$\times 10^{3}$ & -0.6011$\times 10^{3}$ & -0.3035$\times 10^{3}$\\
Hg$^{75+}$ & 5.4463& -0.1262$\times 10^{4}$ & 0.8776$\times 10^{3}$ &  0.1023$\times 10^{4}$ & -0.1246$\times 10^{4}$ & -0.6072$\times 10^{3}$ \\
Bi$^{78+}$ & 5.5211& -0.1562$\times 10^{4}$ & 0.1078$\times 10^{4}$ & 0.1281$\times 10^{4}$ &-0.1541$\times 10^{4}$ & -0.7447$\times 10^{3}$ \\
Fr$^{82+}$ & 5.5915& -0.2076$\times 10^{4}$ & 0.1411$\times 10^{4}$ &  0.1730$\times 10^{4}$ & -0.2042$\times 10^{4}$ & -0.9764$\times 10^{3}$\\
Th$^{85+}$ & 5.7848 & -0.2570$\times 10^{4}$ & 0.1723$\times 10^{4}$ & 0.2171$\times 10^{4}$ & -0.2520$\times 10^{4}$ & -0.1195$\times 10^{4}$\\
U$^{87+}$ & 5.8571 &  -0.2966$\times 10^{4}$ & 0.1968$\times 10^{4}$ & 0.2530$\times 10^{4}$ & -0.2899$\times 10^{4}$ & -0.1368$\times 10^{4}$\\
\end{tabular}
\end{ruledtabular}
\end{table*}

\begin{table*}[H]
\caption{\label{tab3}Mass shift contributions in terms of the $K$-factor
(in units of 1000 GHz$\cdot$amu) for the $2p_{3/2}-2p_{1/2}$
transition in B-like ions. The calculations are performed in the middle basis $(10s\,\,10p\,\,10d\,\,10f\,\,10g)$ of the virtual orbitals, accounting for the single, double and triple excitations.}
\begin{ruledtabular}
\begin{tabular}{cccccccc}
Ions & ${\langle r^2 \rangle}^{1/2}$  & NMS & SMS & RNMS & RSMS & Total \\
O$^{3+}$ & 2.6991&-0.1989$\times 10^{-1}$ &  0.1735$\times 10^{-1}$ & 0.1885$\times 10^{-1}$ & -0.2662$\times 10^{-1}$ & -0.1031$\times 10^{-1}$\\
F$^{4+}$ &2.8976 &-0.3781$\times 10^{-1}$  &0.3206$\times 10^{-1}$  &0.3448$\times 10^{-1}$  &-0.4872$\times 10^{-1}$  &-0.1999$\times 10^{-1}$\\
Ne$^{5+}$ & 3.0055 &-0.6580$\times 10^{-1}$ &  0.5431$\times 10^{-1}$ & 0.5809$\times 10^{-1}$ &  -0.8207$\times 10^{-1}$ & -0.3547$\times 10^{-1}$\\
Na$^{6+}$ &  2.9936 &-0.1071 & 0.8619$\times 10^{-1}$ &  0.9202$\times 10^{-1}$ & -0.1299 & -0.5881$\times 10^{-1}$\\
Al$^{8+}$  & 3.0610 &-0.2446 & 0.1882 & 0.2018 & -0.2846 & -0.1391  \\
P$^{10+}$  & 3.1889 &-0.4857 & 0.3591 & 0.3892 & -0.5476 & -0.2850 \\
S$^{11+}$  & 3.2611 &-0.6582 & 0.4779 & 0.5215 & -0.7328 & -0.3917 \\
Cl$^{12+}$ & 3.3654 &-0.8736 & 0.6234 & 0.6854 & -0.9617 & -0.5266 \\
Ar$^{13+}$ & 3.4028 &-0.1139$\times 10^{1}$ & 0.7993 & 0.8858 & -0.1241$\times 10^{1}$ & -0.6944\\
K$^{14+}$  & 3.4349 &-0.1461$\times 10^{1}$ & 0.1010$\times 10^{1}$ & 0.1128$\times 10^{1}$ & -0.1577$\times 10^{1}$ & -0.9002  \\
Ca$^{15+}$ & 3.4776 & -0.1848$\times 10^{1}$ & 0.1259$\times 10^{1}$ & 0.1418$\times 10^{1}$ & -0.1979$\times 10^{1}$ & -0.1149$\times 10^{1}$\\
Sc$^{16+}$ & 3.4776& -0.2309$\times 10^{1}$ & 0.1553$\times 10^{1}$ & 0.1762$\times 10^{1}$ & -0.2454$\times 10^{1}$ & -0.1448$\times 10^{1}$\\
Ti$^{17+}$ & 3.5921& -0.2853$\times 10^{1}$ &  0.1896$\times 10^{1}$ & 0.2166$\times 10^{1}$ &  -0.3010$\times 10^{1}$ &  -0.1802$\times 10^{1}$\\
V$^{18+}$  & 3.6002&-0.3490$\times 10^{1}$ &  0.2294$\times 10^{1}$ &  0.2639$\times 10^{1}$ & -0.3659$\times 10^{1}$ & -0.2217$\times 10^{1}$ \\
Cr$^{19+}$ & 3.6452& -0.4231$\times 10^{1}$ & 0.2753$\times 10^{1}$ & 0.3187$\times 10^{1}$ & -0.4409$\times 10^{1}$ & -0.2700$\times 10^{1}$\\
Fe$^{21+}$  & 3.7377& -0.6072$\times 10^{1}$ &  0.3887$\times 10^{1}$ & 0.4547$\times 10^{1}$ & -0.6259$\times 10^{1}$ & -0.3896$\times 10^{1}$\\
Co$^{22+}$ & 3.7875 & -0.7197$\times 10^{1}$ & 0.4578$\times 10^{1}$ & 0.5377$\times 10^{1}$ & -0.7382$\times 10^{1}$ & -0.4624$\times 10^{1}$\\
Cu$^{24+}$ & 3.8823 & -0.9928$\times 10^{1}$ & 0.6256$\times 10^{1}$ & 0.7390$\times 10^{1}$ & -0.1009$\times 10^{2}$ &-0.6373$\times 10^{2}$\\
Zn$^{25+}$ &  3.9491 &-0.1156 $\times 10^{2}$ & 0.7265$\times 10^{1}$ & 0.8596$\times 10^{1}$ & -0.1171$\times 10^{2}$ &-0.7409$\times 10^{1}$ \\
Kr$^{31+}$& 4.1835& -0.2630 $\times 10^{2}$ & 0.1654$\times 10^{2}$ &  0.1948$\times 10^{2}$ & -0.2610$\times 10^{2}$ & -0.1638$\times 10^{2}$\\
Mo$^{37+}$ & 4.3151 & -0.5295$\times 10^{2}$ & 0.3406$\times 10^{2}$ &  0.3931$\times 10^{2}$ & -0.5193$\times 10^{2}$ &  -0.3150$\times 10^{2}$\\
Xe$^{49+}$ & 4.7964& -0.1699$\times 10^{3}$ & 0.1155$\times 10^{3}$ &  0.1281$\times 10^{3}$ & -0.1655$\times 10^{3}$ & -0.9175$\times 10^{2}$ \\
Nd$^{55+}$ & 4.9123&  -0.2818$\times 10^{3}$ & 0.1954$\times 10^{3}$ & 0.2152$\times 10^{3}$ & -0.2752$\times 10^{3}$ & -0.1465$\times 10^{3}$\\
Yb$^{65+}$ & 5.0423& -0.6118$\times 10^{3}$ & 0.4296$\times 10^{3}$ &  0.4795$\times 10^{3}$ & -0.6011$\times 10^{3}$ & -0.3038$\times 10^{3}$\\
Hg$^{75+}$ & 5.4463& -0.1264$\times 10^{4}$ & 0.8777$\times 10^{3}$ &  0.1024$\times 10^{4}$ & -0.1246$\times 10^{4}$ & -0.6075$\times 10^{3}$ \\
Bi$^{78+}$ & 5.5211& -0.1564$\times 10^{4}$ & 0.1078$\times 10^{4}$ & 0.1282$\times 10^{4}$ &-0.1541$\times 10^{4}$ & -0.7450$\times 10^{3}$ \\
Fr$^{82+}$ & 5.5915& -0.2078$\times 10^{4}$ & 0.1411$\times 10^{4}$ &  0.1732$\times 10^{4}$ & -0.2042$\times 10^{4}$ & -0.9767$\times 10^{3}$\\
Th$^{85+}$ & 5.7848 & -0.2572$\times 10^{4}$ & 0.1723$\times 10^{4}$ & 0.2173$\times 10^{4}$ & -0.2520$\times 10^{4}$ & -0.1196$\times 10^{4}$\\
U$^{87+}$ & 5.8571 &  -0.2969$\times 10^{4}$ & 0.1968$\times 10^{4}$ & 0.2532$\times 10^{4}$ & -0.2899$\times 10^{4}$ & -0.1368$\times 10^{4}$\\
\end{tabular}
\end{ruledtabular}
\end{table*}

\begin{table*}[H]
\caption{\label{tab4}Mass shift contributions in terms of the $K$-factor (in units of 1000 GHz$\cdot$amu) for the $2p_{3/2}-2p_{1/2}$ transition in B-like ions. The calculations are performed using the CI-DFS method with the large basis $(15s\,\,15p\,\,15d\,\,15f\,\,15g\,\,12f\,\,12g\,\,12h)$ and adding the triple excitation contribution $\Delta_{\rm{triple}}$, which was obtained as the difference between the total values of the $K$-factors from Tables \ref{tab3} and \ref{tab2}.}
\begin{ruledtabular}
\begin{tabular}{cccccccccc}
Ions & ${\langle r^2 \rangle}^{1/2}$ & NMS & SMS & RNMS & RSMS & Total & Total+$\Delta_{\rm triple}$ & C. Naze {\it et al.} {\cite{Naze_2014}}\\
O$^{3+}$ & 2.6991 &-0.1919$\times 10^{-1}$ &  0.1712$\times 10^{-1}$ & 0.1881$\times 10^{-1}$ & -0.2659$\times 10^{-1}$ & -0.0985$\times 10^{-1}$ &  -0.0979$\times 10^{-1}$  &-0.0913$\times 10^{-1}$ \\
F$^{4+}$ & 2.8976 &-0.3694$\times 10^{-1}$  &0.3180$\times 10^{-1}$  & 0.3443$\times 10^{-1}$  &-0.4867$\times 10^{-1}$  &-0.1939$\times 10^{-1}$  
&-0.1965$\times 10^{-1}$ & -0.2130$\times 10^{-1}$ \\
Ne$^{5+}$ & 3.0055 &-0.6472$\times 10^{-1}$ &   0.5403$\times 10^{-1}$ & 0.5802$\times 10^{-1}$ & -0.8202$\times 10^{-1}$ & -0.3468$\times 10^{-1}$ &  -0.3458$\times 10^{-1}$  & -0.3411$\times 10^{-1}$ \\
Na$^{6+}$ &  2.9936 &-0.1057 & 0.8588$\times 10^{-1}$ & 0.9194$\times 10^{-1}$ & -0.1299 & -0.5779$\times 10^{-1}$ & -0.5769$\times 10^{-1}$ &-0.5687$\times 10^{-1}$\\
Al$^{8+}$ & 3.0610 & -0.2424 &0.1878 & 0.2017   &-0.2845  &-0.1375 & -0.1374 &-0.1396\\
P$^{10+}$ & 3.1889 &-0.4822& 0.3584 &   0.3891 &  -0.5477 &  -0.2825 &-0.2825 &-0.2811\\
S$^{11+}$ & 3.2611 &-0.6540  &  0.4770& 0.5213 &-0.7329 & -0.3886& -0.3887&-0.3872\\
Cl$^{12+}$ & 3.3654&-0.8683 &  0.6222 & 0.6852& -0.9618 & -0.5227 & -0.5230&-0.5251\\
Ar$^{13+}$ & 3.4028& -0.1132$\times 10^1$ & 0.7976 & 0.8855 & -0.1241$\times 10^1$ & -0.6900 & -0.6905 &
-0.6888\\
K$^{14+}$ & 3.4349 & -0.1453$\times 10^1$ &  0.1008$\times 10^1$ & 0.1128$\times 10^1$ & -0.1578$\times 10^1$ & -0.8946 & -0.8953 & -0.8940 \\
Ca$^{15+}$ &  3.4776 & -0.1838$\times 10^1$ & 0.1257$\times 10^1$ & 0.1417$\times 10^1$ & -0.1979$\times 10^1$ & -0.1143$\times 10^1$ & -0.1144$\times 10^1$
&-0.1143$\times 10^1$\\
Sc$^{16+}$ &  3.4776 & -0.2297$\times 10^1$ &0.1550$\times 10^1$ & 0.1761$\times 10^1$ & -0.2454$\times 10^1$ &-0.1440
$\times 10^1$ & -0.1442$\times 10^1$&
-0.1441$\times 10^1$\\
Ti$^{17+}$ & 3.5921 & -0.2838$\times 10^1$ &  0.1892$\times 10^1$ & 0.2165$\times 10^1$  & -0.3011$\times 10^1$ & -0.1792$\times 10^1$ & -0.1795$\times 10^1$  
&-0.1795$\times 10^1$ \\
V$^{18+}$  & 3.6002&-0.3472$\times 10^1$ & 0.2289$\times 10^1$ &0.2637$\times 10^1$ & -0.3659$\times 10^1$ &  -0.2205$\times 10^1$ & -0.2208$\times 10^1$&-0.2210$\times 10^1$ \\
Cr$^{19+}$ & 3.6452&-0.4210$\times 10^1$  & 0.2748$\times 10^1$ & 0.3185$\times 10^1$ & -0.4410$\times 10^1$ & -0.2686$\times 10^1$& -0.2690$\times 10^1$&-0.2694$\times 10^1$\\
Fe$^{21+}$  & 3.7377&-0.6042$\times 10^1$ & 0.3880$\times 10^1$ & 0.4544$\times 10^1$ & -0.6259$\times 10^1$ & -0.3878$\times 10^1$& -0.3884$\times 10^1$&-0.3895$\times 10^1$\\
Co$^{22+}$  &3.7875 & -0.7163$\times 10^1$ &0.4570$\times 10^1$  &0.5373$\times 10^1$  &-0.7383$\times 10^1$ &-0.4602$\times 10^1$ & -0.4609$\times 10^1$ &-0.4626$\times 10^1$\\
Cu$^{24+}$ & 3.8823 &-0.9882$\times 10^1$& 0.6245$\times 10^1$ &0.7384$\times 10^1$ &-0.1009$\times 10^1$ &-0.6344$\times 10^1$ &
 -0.6363$\times 10^1$ &-0.6385$\times 10^1$\\
Zn$^{25+}$ & 3.9491 &-0.1151$\times 10^2$ & 0.7252$\times 10^1$ & 0.8589$\times 10^1$ & -0.1171$\times 10^2$ & -0.7376$\times 10^1$ & 
 -0.7389$\times 10^1$&-0.7429$\times 10^1$\\
Kr$^{31+}$& 4.1835 &-0.2619$\times 10^2$ & 0.1651$\times 10^2$ &  0.1946$\times 10^2$ & -0.2610$\times 10^2$ & -0.1632$\times 10^2$ & -0.1635$\times 10^2$& -0.1649$\times 10^2$ \\
Mo$^{37+}$ & 4.3151& -0.5275$\times 10^2$ &  0.3402$\times 10^2$ &  0.3926$\times 10^2$ & -0.5193$\times 10^2$ & -0.3139$\times 10^2$ & -0.3145$\times 10^2$ & -0.3180$\times 10^2$ \\
Xe$^{49+}$ &4.7964 &-0.1694$\times 10^3$ & 0.1155$\times 10^3$ & 0.1279$\times 10^3$ & -0.1655$\times 10^3$ & 
 -0.9151$\times 10^2$ & -0.9165$\times 10^2$ &-\\
Nd$^{55+}$ & 4.9123&-0.2811$\times 10^3$ & 0.1953$\times 10^3$ & 0.2148$\times 10^3$ & -0.2752$\times 10^3$   
 &-0.1462$\times 10^3$ & -0.1464$\times 10^3$&-\\
Yb$^{65+}$ & 5.0423&-0.6106$\times 10^3$ & 0.4295$\times 10^3$&0.4788$\times 10^3$ & -0.6011$\times 10^3$ & -0.3033$\times 10^3$ & -0.3036$\times 10^3$&-\\
Hg$^{75+}$ & 5.4463& -0.1262$\times 10^4$ & 0.8776$\times 10^3$  & 0.1023$\times 10^4$ & -0.1246$
\times 10^4$ & -0.6069$\times 10^3$ & -0.6071$\times 10^3$&-\\
Bi$^{78+}$ & 5.5211& -0.1562$\times 10^4$ & 0.1078$\times 10^4$ & 0.1281$\times 10^4$ & -0.1541$\times 10^4$ & -0.7444$\times 10^3$ & -0.7447$\times 10^3$&-\\
Fr$^{82+}$ & 5.5915&-0.2075$\times 10^4$ & 0.1411$\times 10^4$ & 0.1730$\times 10^4$ & -0.2042$\times 10^4$ & -0.9761$\times 10^3$ & -0.9764$\times 10^3$& -\\
Th$^{85+}$ & 5.7848& -0.2569$\times 10^4$ & 0.1723$\times 10^4$ & 0.2170$\times 10^4$ & -0.2520$\times 10^4$&  -0.1195 $\times 10^4$ & -0.1196$\times 10^4$ &-\\
U$^{87+}$ & 5.8571&-0.2965$\times 10^4$ & 0.1968$\times 10^4$ & 0.2529$\times 10^4$ & -0.2899$\times 10^4$ & -0.1368$\times 10^4$ &-0.1368$\times 10^4$ &-\\
\end{tabular}
\end{ruledtabular}
\end{table*}

\begin{table*}[H]
\caption{\label{tab5}The results of the numerical calculations of the two-electron nuclear recoil contribution  for
the $(1s)^2 2p_{1/2}$ and $(1s)^2 2p_{3/2}$ states  of lithiumlike ions, expressed in terms of the function $Q(\alpha Z)$ defined by equation (\ref{2QED}).}
\begin{ruledtabular}
\begin{tabular}{ccc|ccc}
\multicolumn{3}{c|}{$(1s)^2 2p_{1/2}$ } & \multicolumn{2}{c}{$(1s)^2 2p_{3/2}$ }  \\
Ion & point nucleus & extended nucleus & point nucleus & extended nucleus   \\
\hline
8  & 0.99834  & 0.99834   & 0.99879 & 0.99879  \\ 
9  & 0.99790 & 0.99790 &  0.99847  &  0.99847 \\
10 & 0.99741&  0.99741 &  0.99811 & 0.99811  \\
11 & 0.99686& 0.99686  & 0.99771 & 0.99771    \\
13 & 0.99561 &  0.99561 & 0.99681  & 0.99681\\
15 &  0.99416 &  0.99416  &  0.99576  &  0.99576\\
16 & 0.99335  &  0.99335  &  0.99517  & 0.99517  \\
17 & 0.99249 & 0.99249   &   0.99455 &  0.99456 \\
18 & 0.99158 & 0.99158   &  0.99389&  0.99390 \\
19 & 0.99061 &0.99061  &  0.99321  &   0.99321\\
20 & 0.98959& 0.98960  & 0.99248  & 0.99248 \\
21 & 0.98852 & 0.98852   &  0.99171  &  0.99172  \\
22 & 0.98740 & 0.98740  &  0.99091  & 0.99092   \\
23 & 0.98622 &0.98622  & 0.99008 &  0.99008\\
24 & 0.98499 &0.98499 &  0.98921 & 0.98921  \\
26 &  0.98236 & 0.98236  & 0.98736  & 0.98736\\
28 & 0.97951 & 0.97952  &  0.98536  & 0.98537\\
29 &  0.97801 &0.97801   &  0.98432  &  0.98432  \\
30 &  0.97645 &  0.97645 & 0.98323 &  0.98324 \\
36 & 0.96592& 0.96593  & 0.97603  &  0.97603 \\
40 & 0.95776 &0.95776   & 0.97056 & 0.97057  \\
42 &  0.95332 & 0.95333  & 0.96763 & 0.96764 \\
54 & 0.92149 & 0.92152 &   0.94742 & 0.94746  \\
60 &  0.90195 & 0.90198 &  0.93567 & 0.93574 \\
70 & 0.86320 & 0.86327 & 0.91367   &  0.91384  \\
80 &  0.81529     & 0.81541 &  0.88839  &  0.88879\\
83 &  0.79879&  0.79893 & 0.88008 & 0.88059\\
87 & 0.77501 & 0.77518 & 0.86838  & 0.86913   \\
90 &  0.75570& 0.75590 &  0.85908   &  0.86007 \\
92 &  0.74206& 0.74229 & 0.85261& 0.85380\\
\end{tabular}
\end{ruledtabular}
\end{table*}

\begin{table*}[H]
\caption{\label{tab6}Total mass shifts in terms of the $K$-factor
(in units of 1000 GHz$\cdot$amu and in units of eV$\cdot$amu) for the $2p_{3/2}-2p_{1/2}$ 
transition in B-like ions. The QED corrections have been evaluated for the Coulomb (${\rm{QED}}_{\rm{Coul}}$) potential and for the effective (${\rm{QED}}_{\rm{LDF}}$) potential, which partly accounts for the screening effects.}
\begin{ruledtabular}
\begin{tabular}{cccccccc}
Ion & ${\langle r^2 \rangle}^{1/2}$ & Total non-QED MS & ${\rm{QED}}_{\rm{Coul}}$ & ${\rm{QED}}_{\rm{LDF}}$ &  \multicolumn{2}{c}{Total MS with QED} \\ 
&&&&& 1000 GHz$\cdot$amu & eV$\cdot$amu\\
O$^{3+}$   & 2.6991 & -9.79$\times 10^{-3}$ &  0.08$\times 10^{-3}$ & 0.03$\times 10^{-3}$ & -9.76(33)$\times 10^{-3}$ & -0.0404(14)$\times 10^{-3}$  \\ 
F$^{4+}$   & 2.8976 & -1.96$\times 10^{-2}$ &0.02$\times 10^{-2}$ &   0.01$\times 10^{-2}$ & -1.96(8)$\times 10^{-2}$  & -0.0081(3)$\times 10^{-2}$ \\
Ne$^{5+}$  & 3.0055 & -3.468$\times 10^{-2}$ &0.031$\times 10^{-2}$& 0.013$\times 10^{-2}$ & -3.45(3)$\times 10^{-2}$ & -0.01425(12)$\times 10^{-2}$ \\
Na$^{6+}$  & 2.9936 & -5.77$\times 10^{-2}$ & 0.06$\times 10^{-2}$& 0.03$\times 10^{-2}$ & -5.74(3)$\times 10^{-2}$ & -0.02375(13)$\times 10^{-2}$\\
Al$^{8+}$  & 3.0610 & -1.375$\times 10^{-1}$ & 0.015$\times 10^{-1}$ & 0.008$\times 10^{-1}$  & -1.366(8)$\times 10^{-1}$ & -0.00565(3)$\times 10^{-1}$ \\
P$^{10+}$  & 3.1889 & -2.825$\times 10^{-1}$ &  0.034$\times 10^{-1}$& 0.020$\times 10^{-1}$ & -2.805(18)$\times 10^{-1}$ & -0.01160(8)$\times 10^{-1}$ \\
S$^{11+}$  & 3.2847 & -3.889$\times 10^{-1}$  & 0.050$\times 10^{-1}$&  0.030$\times 10^{-1}$ & -3.857(21)$\times 10^{-1}$ & -0.01595(9)$\times 10^{-1}$\\
Cl$^{12+}$ & 3.3840 & -5.227$\times 10^{-1}$  &0.071$\times 10^{-1}$& 0.044$\times 10^{-1}$ & -5.186(28)$\times 10^{-1}$ & -0.02145(12)$\times 10^{-1}$\\
Ar$^{13+}$ & 3.4028 & -6.90$\times 10^{-1}$  & 0.10$\times 10^{-1}$& 0.06$\times 10^{-1}$ & -6.84(4)$\times 10^{-1}$ &-0.02829(15)$\times 10^{-1}$ \\
K$^{14+}$  & 3.4349 & -8.95$\times 10^{-1}$  &  0.14$\times 10^{-1}$&  0.09$\times 10^{-1}$ & -8.86(5)$\times 10^{-1}$ & -0.0367(2)$\times 10^{-1}$ \\
Ca$^{15+}$ & 3.4776 & -1.144                 & 0.018& 0.012 & -1.131(6) & -0.00468(3)\\
Sc$^{16+}$ & 3.5459 & -1.441   &  0.025& 0.017 & -1.424(9) & -0.00589(3)\\
Ti$^{17+}$ & 3.5921 & -1.794 & 0.032& 0.023 & -1.771(11) & -0.00732(4)\\
V$^{18+}$  & 3.6002 & -2.208 &  0.042 & 0.030 & -2.178(13) & -0.00901(6)\\
Cr$^{19+}$ & 3.6452 & -2.690 &  0.054 & 0.039 & -2.650(17) & -0.01096(7)\\
Fe$^{21+}$ & 3.7377 & -3.884 & 0.086& 0.065 & -3.819(25) & -0.01579(10)\\
Co$^{22+}$ & 3.7875 & -4.61 & 0.13&  0.11 & -4.51(3) & -0.01864(14)\\
Cu$^{24+}$ & 3.9022 & -6.36 & 0.16 & 0.13 & -6.24(5) & -0.0258(2)\\
Zn$^{25+}$ & 3.9491 & -7.39 &  0.20&  0.15 & -7.23(5) & -0.0299(2)\\
Kr$^{31+}$ & 4.1835 & -1.635$\times 10^{1}$ &   0.058$\times 10^{1}$ & 0.048$\times 10^{1}$ & -1.587(12)$\times 10^{1}$ & -0.00656(5)$\times 10^{1}$\\
Mo$^{37+}$ & 4.3151 & -3.145$\times 10^{1}$ & 0.145$\times 10^{1}$& 0.124$\times 10^{1}$ & -3.021(25)$\times 10^{1}$ & -0.01250(10)$\times 10^{1}$ \\
Xe$^{49+}$ & 4.7964 & -9.16$\times 10^{1}$  &  0.65$\times 10^{1}$& 0.58$\times 10^{1}$ & -8.59(8)$\times 10^{1}$ & -0.0355(3)$\times 10^{1}$ \\
Nd$^{55+}$ & 4.9123 & -1.464$\times 10^{2}$  & 0.121$\times 10^{2}$& 0.109$\times 10^{2}$ & -1.354(13)$\times 10^{2}$ & -0.00560(5)$\times 10^{2}$ \\
Yb$^{65+}$ & 5.3215 & -3.036$\times 10^{2}$  & 0.030$\times 10^{3}$& 0.028$\times 10^{3}$ & -2.757(26)$\times 10^{2}$ & -0.01140(11)$\times 10^{2}$\\
Hg$^{75+}$ & 5.4463 & -0.607$\times 10^{3}$  &0.064$\times 10^{3}$ & 0.061$\times 10^{3}$ & -0.547(4)$\times 10^{3}$ & -0.00226(2)$\times 10^{3}$ \\
Bi$^{78+}$ & 5.5211 & -0.745$\times 10^{3}$ & 0.078$\times 10^{3}$& 0.074$\times 10^{3}$ & -0.671(5)$\times 10^{3}$ & -0.00277(2)$\times 10^{3}$\\
Fr$^{82+}$ & 5.5915 & -0.976$\times 10^3$ & 0.098$\times 10^3$&  0.094$\times 10^3$ & -0.883(6)$\times 10^3$ & -0.00365(2)$\times 10^3$ \\
Th$^{85+}$ &5.7848  & -1.195$\times 10^3$& 0.112$\times 10^3$& 0.109$\times 10^3$ & -1.086(8)$\times 10^3$ & -0.00449(3)$\times 10^3$ \\
U$^{87+}$  & 5.8571 & -1.368$\times 10^3$ & 0.121$\times 10^3$& 0.118$\times 10^3$ & -1.250(12)$\times 10^3$ & -0.00517(5)$\times 10^3$ \\
\end{tabular}
\end{ruledtabular}
\end{table*}

\begin{table*}[H]
\caption{\label{tab7} Field shifts in terms of the $F$-factor
(in MHz/$\rm{{fm}^2}$ and in meV/$\rm{{fm}^2}$) for the $2p_{3/2}-2p_{1/2}$ 
transition in B-like ions.}
\begin{ruledtabular}
\begin{tabular}{cccccc}
Ion & ${\langle r^2 \rangle}^{1/2}$ & DF &  \multicolumn{2}{c}{CI-DFS+Breit}  & C. Naze {\it et al.} \cite{Naze_2014}\\
&&&MHz/$\rm{{fm}^2}$ & meV/$\rm{{fm}^2}$&MHz/$\rm{{fm}^2}$\\
\hline
O$^{3+}$   & 2.6991 &  0.133  &   0.446(50)$\times 10^{-1}$  &  0.184(2)$\times 10^{-6}$ 
 & 0.5$\times 10^{-1}$ \\
F$^{4+}$   & 2.8976 &  0.214  &  -0.302(40)$\times 10^{-2}$ &  -0.125(16)$\times 10^{-7}$ & 0  \\
Ne$^{5+}$  & 3.0055 &  0.294  & -0.197(25)$\times 10^{0}$& -0.81(10)$\times 10^{-6}$   & -0.22\\
Na$^{6+}$  & 2.9936 &  0.322  & -0.703(88)$\times 10^{0}$ & -0.291(36)$\times 10^{-5}$ & -0.75\\
Al$^{8+}$  & 3.1224 & -0.131  & -0.389(45)$\times 10^{1}$  &-0.161(19)$\times 10^{-4}$
&-0.407$\times 10^{1}$  \\
P$^{10+}$  & 3.1889 & -0.247$\times 10^{1}$ &  -0.138(15)$\times 10^{2}$  & -0.57(6)$\times 10^{-4}$& -0.144$\times 10^{2}$ \\
S$^{11+}$  & 3.2847 & -0.512$\times 10^{1}$ & -0.238(25)$\times 10^{2}$  & -0.98(10)$\times 10^{-4}$& -0.248$\times 10^{2}$ \\
Cl$^{12+}$ & 3.3840 & -0.942$\times 10^{1}$ &  -0.390(40)$\times 10^{2}$ & 
-1.61(17)$\times 10^{-4}$&-0.406$\times 10^{2}$\\
Ar$^{13+}$ & 3.4028 & -0.160$\times 10^{2}$ & -0.617(50)$\times 10^{2}$ & 
-2.55(21)$\times 10^{-3}$& -0.641$\times 10^{2}$
\\
K$^{14+}$  & 3.4349 & -0.258$\times 10^{2}$ & -0.945(70)$\times 10^{2}$  & 
-0.39(29)$\times 10^{-3}$&-0.982$\times 10^{2}$ 
\\
Ca$^{15+}$ & 3.4776 & -0.400$\times 10^{2}$ & -0.141(10)$\times 10^{3}$  & 
-0.58(4)$\times 10^{-3}$&-0.1463$\times 10^{3}$ \\
Sc$^{16+}$ & 3.5459 & -0.599$\times 10^{2}$ & -0.205(14)$\times 10^{3}$ & 
-0.85(6)$\times 10^{-3}$ &-0.2126$\times 10^{3}$ \\
Ti$^{17+}$ & 3.5921 & -0.873$\times 10^{2}$ & -0.292(18)$\times 10^{3}$ & 
-1.21(7)$\times 10^{-3}$&-0.3028$\times 10^{3}$ 
\\
V$^{18+}$  & 3.6002 & -0.124$\times 10^{3}$ & -0.408(21)$\times 10^{3}$ & 
-1.69(9)$\times 10^{-3}$&-0.423$\times 10^{3}$ \\
Cr$^{19+}$ & 3.6452 & -0.174$\times 10^{3}$ & -0.561(25)$\times 10^{3}$ & -2.32(10)$\times 10^{-3}$ &-0.582$\times 10^{3}$ \\
Fe$^{21+}$ & 3.7377 & -0.324$\times 10^{3}$ & -0.101(4)$\times 10^{4}$  & -4.18(16)$\times 10^{-3}$&-0.1051$\times 10^{3}$ \\
Co$^{22+}$ & 3.7875 & -0.433$\times 10^{3}$ & -0.133(5)$\times 10^{4}$ &-5.50(21)$\times 10^{-3}$ &-0.1385$\times 10^{4}$ \\
Cu$^{24+}$ & 3.9022 & -0.747$\times 10^{3}$ & -0.223(6)$\times 10^{4}$ & -9.22(25)$\times 10^{-3}$&-0.2326$\times 10^{4}$\\
Zn$^{25+}$ & 3.9491 & -0.967$\times 10^{3}$ & -0.285(6)$\times 10^{4}$ & -1.18(26)$\times 10^{-2}$&-0.2969$\times 10^{4}$ \\
Kr$^{31+}$ & 4.1835 & -0.387$\times 10^{4}$ & -0.103(2)$\times 10^{5}$ &  -0.426(8)$\times 10^{-1}$&-0.1076$\times 10^{5}$ \\
Mo$^{37+}$ & 4.3151 & -0.127$\times 10^{5}$ & -0.300(5)$\times 10^{5}$ & -1.241(21)$\times 10^{-1}$&-0.316$\times 10^{5}$\\
Xe$^{49+}$ & 4.7964 & -0.967$\times 10^{5}$ & -0.176(3)$\times 10^{6}$ &-7.28(12)$\times 10^{-1}$&-\\
Nd$^{55+}$ & 4.9123 & -0.240$\times 10^{6}$ & -0.388(5)$\times 10^{6}$ & -1.605(21)& -\\
Yb$^{65+}$ & 5.3215 & -0.996$\times 10^{6}$ & -0.137(2)$\times 10^{7}$ & -5.67(8)&-\\
Hg$^{75+}$ & 5.4463 & -0.380$\times 10^{7}$ & -0.468(6)$\times 10^{7}$ & -1.935(25)$\times 10^{1}$&-\\
Bi$^{78+}$ & 5.5211 & -0.564$\times 10^{7}$ & -0.678(9)$\times 10^{7}$ & -2.804(36)$\times 10^{1}$ &-\\
Fr$^{82+}$ & 5.5915 & -0.958$\times 10^{7}$ & -0.112(2)$\times 10^{8}$ & -4.63(6)$\times 10^{1}$& -\\
Th$^{85+}$ & 5.7848  & -0.140$\times 10^{8}$ & -0.160(2)$\times 10^{8}$ &-6.62(9)$\times 10^{1}$ &-\\
U$^{87+}$  & 5.8571  & -0.182$\times 10^{8}$ & -0.206(3)$\times 10^{8}$ & -8.52(11)$\times 10^{1}$& -\\
\end{tabular}
\end{ruledtabular}
\end{table*}

\begin{table*}[H]
\caption{\label{tab8}QED corrections to the field-shift in terms of the $F$-factor
(in MHz/$\rm{{fm}^2}$) for the $2p_{3/2}-2s$ and $2p_{1/2}-2s$ transitions in high-$Z$ Li-like ions, and also for the $2p_{3/2}-2p_{1/2}$ transition in high-$Z$ B-like ions.}
\begin{ruledtabular}
\begin{tabular}{ccc|cc|cc}
& \multicolumn{2}{c|}{$2p_{1/2}-2s$} & \multicolumn{2}{c|}{$2p_{3/2}-2s$} & 
$2p_{3/2}-2p_{1/2}$ \\
Ion & this work &  Ref. {\cite{Zubova_2014}} & this work & Ref. {\cite{Zubova_2014}} & this work\\
Bi$^{78+}$ & 0.0439(35)$\times 10^{7}$   & 0.039(11)$\times 10^{7}$ & 0.0448(36)$\times 10^{7}$  & 0.051(14)$\times 10^{7}$
&0.0063(21)$\times 10^{6}$  \\
Fr$^{82+}$ & 0.0645(50)$\times 10^{7}$    &  0.056(18) $\times 10^{7}$&0.0659(53)$\times 10^{7}$   & 0.078(25)$\times 10^{7}$ & 0.0112(32)$\times 10^{6}$ \\
Th$^{85+}$ & 0.0853(68)$\times 10^{7}$    &  0.073(25)$\times 10^{7}$  &0.0876(70)$\times 10^{7}$   &   0.107(34)$\times 10^{7}$  &0.0173(40)$\times 10^{6}$  \\
U$^{87+}$  & 0.1026(82)$\times 10^{7}$    & 0.087(30)$\times 10^{7}$  &
0.1055(85)$\times 10^{7}$   & 0.132(43)$\times 10^{7}$ & 0.0230(45)$\times 10^{6}$ & \\
\end{tabular}
\end{ruledtabular}
\end{table*}

\begin{table*}[H]
\caption{\label{tab9} Individual contributions to the isotope shift for the $2p_{3/2}-2p_{1/2}$ transition in B-like
$^{40,36}\rm{Ar}^{13+}$ (in $cm^{-1}$) with given values of $\delta {\langle r^2 \rangle}$=0.251 $\rm{fm^2}$ \cite{Angeli_2013}.} 

\begin{ruledtabular}
\begin{tabular}{cc|}
{ {Main contributions}}&\\
Field shift  &  -0.0005\\
Mass shift   &  0.0640\\
\hline
FS plus MS (this work) & 0.0635\\
FS plus MS  (I. I. Tupitsyn {\it {et al.}} \cite{Tupitsyn_2003}) & 0.0635 \\
FS plus MS (C. Naze {\it {et al.}} \cite{Naze_2014})) & 0.0633\\
{${\rm{QED}}_{\rm{LDF}}$}&\\ 
Mass shift (this work) & -0.0006\\
Mass shift (R. Soria Orts {\it {et al.}} \cite{Orts_2006}) & -0.0006\\
\hline
Total IS theory (this work) \footnotemark  & 0.0629(3)\\
Total IS theory (R. Soria Orts {\it {et al.}} \cite{Orts_2006}) & 0.0629\\
Total IS experiment (R. Soria Orts {\it {et al.}} \cite{Orts_2006}) & 0.0629\\
\end{tabular}
\footnotetext{The uncertainty of  $\delta{\langle r^2 \rangle}$ is not included.}
\end{ruledtabular}
\end{table*}

\begin{table*}[H]
\caption{\label{tab10} Individual contributions to the isotope shifts for the $2p_{3/2}-2p_{1/2}$ transition in B-like 
 $^{238,236}\rm{U}^{87+}$, $^{238,234}\rm{U}^{87+}$ (in meV) with given values of $\delta {\langle r^2 \rangle}$. The values of $\delta {\langle r^2 \rangle}$ are taken from Ref. \cite{Angeli_2013}.}
\begin{ruledtabular}
\begin{tabular}{cc|c}
& $^{238,236}\rm{U}^{87+}$& $^{238,234}\rm{U}^{87+}$\\[1mm]
 & {$^{238,236}\delta {\langle r^2 \rangle}$=0.1676 $\rm{fm^2}$}& {$^{238,234}\delta{\langle r^2 \rangle}$=0.334 $\rm{fm^2}$}\\[1mm] 
\hline
{ {Main contributions}}&&\\
Field shift  & -14.26   & -28.41 \\
Mass shift   & 0.20 & 0.41\\
\hline
{ QED}&&\\
Field shift  & 0.02 & 0.03\\
Mass shift & -0.07 & -0.14 \\
Nuclear polarization & 0.16 & 0.30\\

Nuclear deformation & -0.1 & -0.2 \\
\hline
Total IS theory (this work) \footnotemark  & -14.1(4) & -28.0(5)\\
\end{tabular}
\footnotetext{The uncertainty of  $\delta{\langle r^2 \rangle}$ is not included.}

\end{ruledtabular}
\end{table*}
\endgroup

\begin{figure}[h]
\center{\includegraphics[width=0.5\linewidth]{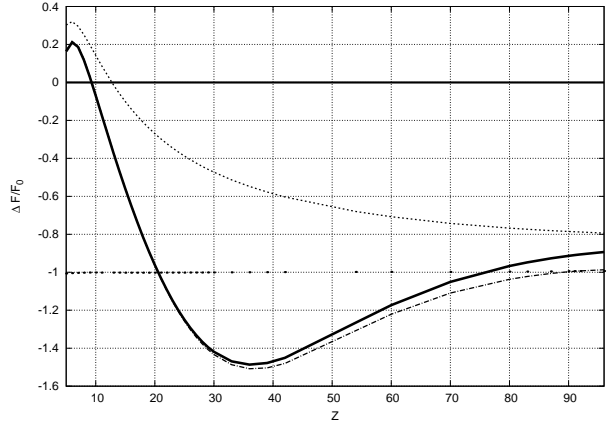}}
\caption{The normalized $F$-factor, $\Delta F/F_0$, where $\Delta F=F_{2p_{3/2}}-F_{2p_{1/2}}$ and $F_0$ is the field shift factor for the $2p_{1/2}$ state of
the hydrogenlike ion, defined by
Eq. ({\ref{F0}}). The dotted line represents the results for the H-like ions obtained using Eq.(\ref{FS_2}), the dashed line shows the results of the DF calculations using Eq. (\ref{FS_2}), the dashed-dotted line indicates the CI-DFS calculations using the approximate formula (\ref{FS_3}), and the solid line represents the results of the CI-DFS calculations using Eq.(\ref{FS_2}).} 
\label{FS_graph1}
\end{figure}

\begin{acknowledgments}
This work was supported by RFBR (Grants No. 13-02-00630 and No. 16-02-00334), SPbSU (Grants No. 11.38.269.2014,  No. 11.42.1478.2015, 11.38.237.2015), and FAIR-Russia
Research Centre (FRRC).  N.A.Z. and A.V.M. acknowledge the financial support 
by the Dynasty 
foundation, G-RISC, and DAAD. 
\end{acknowledgments}


\end{document}